\documentclass[11pt]{article}

\usepackage{times}
\usepackage{epsfig}

\title{Scaling laws in the functional content of genomes}

\author{Erik van Nimwegen\\
\\
\normalsize{Center for Studies in Physics and Biology, the Rockefeller University,}\\
\normalsize{1230 York Avenue, New York, NY 12001, USA}\\
\\
\normalsize{E-mail:  erik@golem.rockefeller.edu.}
}
\date{}
\begin{document}

\maketitle 

{\bf With the number of sequenced genomes now over one hundred, and
the availability of rough functional annotations for a substantial
proportion of their genes, it has become possible to study the
statistics of gene content across genomes. Here I show that, for many
high-level functional categories, the number of genes in the category
scales as a power-law in the total number of genes in the genome. The
occurrence of such scaling laws can be explained with a simple
theoretical model, and this model suggests that the exponents of the
observed scaling laws correspond to universal constants of the
evolutionary process. I discuss some consequences of these
scaling laws for our understanding of organism design.}

\vskip0.5cm

What fraction of a genome's gene content is allotted to different
functional tasks, and how does this depend on the complexity of the
organism? Until recently, there was simply no data to address such
questions in a quantitative way. Presently, however, there are more
than $100$ sequenced genomes \cite{ncbi_genomes} in public databases,
and protein-family classification algorithms allow functional
annotations for a considerable fraction of the genes in each
genome. Thus, it has become possible to analyze the statistics of
functional gene-content {\em across} different genomes, and here I
present results on the dependency of the number of genes in different
high-level categories on the total number of genes in the genome.

\section*{Evaluating the functional gene-content of genomes}

To estimate the number of genes in different functional categories
each genome has to be functionally annotated. The main results
presented in this paper were obtained using the Interpro
\cite{interpro_ref} annotations of sequenced genomes available from
the European Bioinformatics Institute \cite{proteome_database}. To map
the Interpro annotations to high-level functional categories I used
the Gene Ontology ``biological process'' hierarchy
\cite{geneontologypaper} and a mapping from Interpro entries to
GO-categories both of which can be obtained from the gene ontology
website \cite{GOwebsite}. For each GO category I collect all Interpro
entries that map to it or to one of its descendants in the
``biological process'' hierarchy. To minimize the effects of potential
biases in the mappings from Interpro to GO I only use high-level
functional categories that are represented by at least $50$ different
Interpro entries. This leaves $44$ high-level GO categories.

A gene with multiple hits to Interpro entries that are associated with
a GO category has a higher probability to belong to that category than
a gene with only a single hit. To take this information into account,
I assume that if a gene $i$ has $n^c_i$ independent hits to Interpro
entries associated with GO category $c$, than with probability
$1-\exp(-\beta n^c_i)$ the gene belongs to the GO category. The
results are generally insensitive to the value of $\beta$ (see the
appendix), and I used $\beta = 3$ for the results shown
below. The estimated number of genes $n_c$ in a genome for a given GO
category is then the sum $n_c = \sum_{i} 1-\exp(-\beta n^c_i)$ over
all genes $i$ in the genome.

\begin{figure}[htbp]
\centerline{\epsfig{file=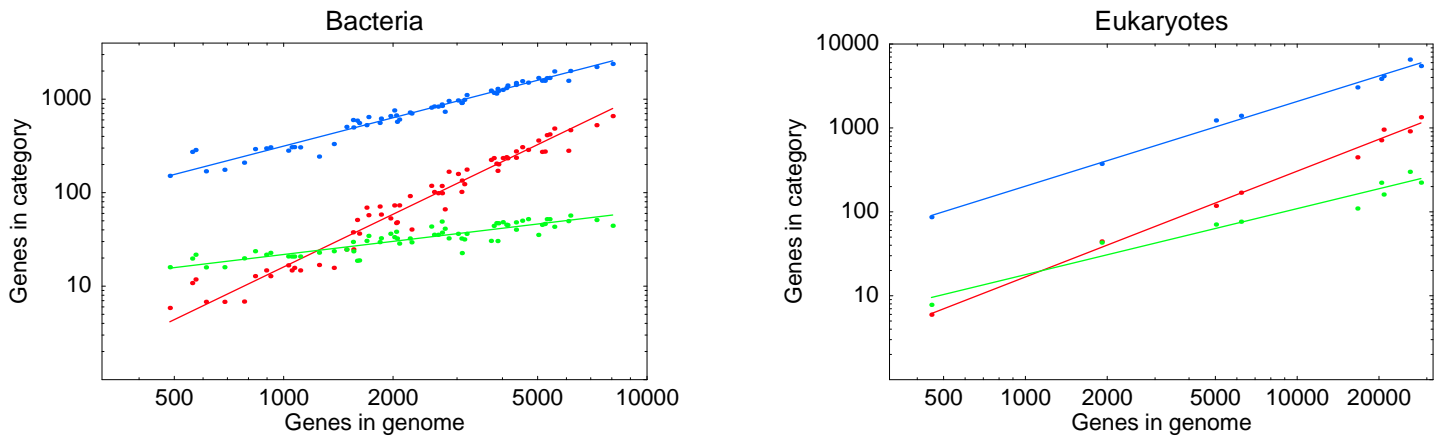,height=6cm}}
\caption{The number of transcription regulatory genes (red), metabolic
genes (blue), and cell cycle related genes (green) as a function of
the total number of genes in the genome for bacteria ({\bf A}) and
eukaryotes ({\bf B}). Both axes are shown on logarithmic scales. Each
dot corresponds to a genome. The straight lines are power-law
fits. Archaea are not shown because the range of genome sizes in
archaea is too small for meaningful fits. For completeness the
archaeal results are shown in figure \ref{archaea_fig} in the
appendix.}
\label{scaling_fig}
\end{figure}

Figure \ref{scaling_fig} shows the results for the categories of
transcription regulatory genes (red), metabolic genes (blue), and
cell-cycle related genes (green) for bacteria (panel A) and eukaryotes
(panel B). Remarkably, for each functional category shown, we find an
approximately power-law relationship (solid line fits)\footnote{These
power-laws as a function of total gene number should be distinguished
from the power-law distributions of gene-family sizes and other
genomic attributes within a {\em single} genome
\cite{Huynen&vanNimwegen1998,LuscombeEtAl2002}.}. That is, if $n_c$ is
the number of genes in the category, and $g$ is the total number of
genes in the genome, we observe laws of the form: $n_c = \lambda
g^{\alpha}$, where both $\lambda$ and the exponent $\alpha$ depend on
the category under study. In fact, such power-laws are observed for
most of the $44$ high-level categories, and the estimated values of
the exponents for several functional categories are shown in Table
\ref{estimated_exponent_table}. Note that a potential source of bias
in estimating these exponents is the occurrence of multiple genomes
from the same or closely-related species in the data. As shown in the
appendix, removing this ``redundancy'' from the data does not alter
the observed exponents.

\begin{table}[htbp]
\centerline{\epsfig{file=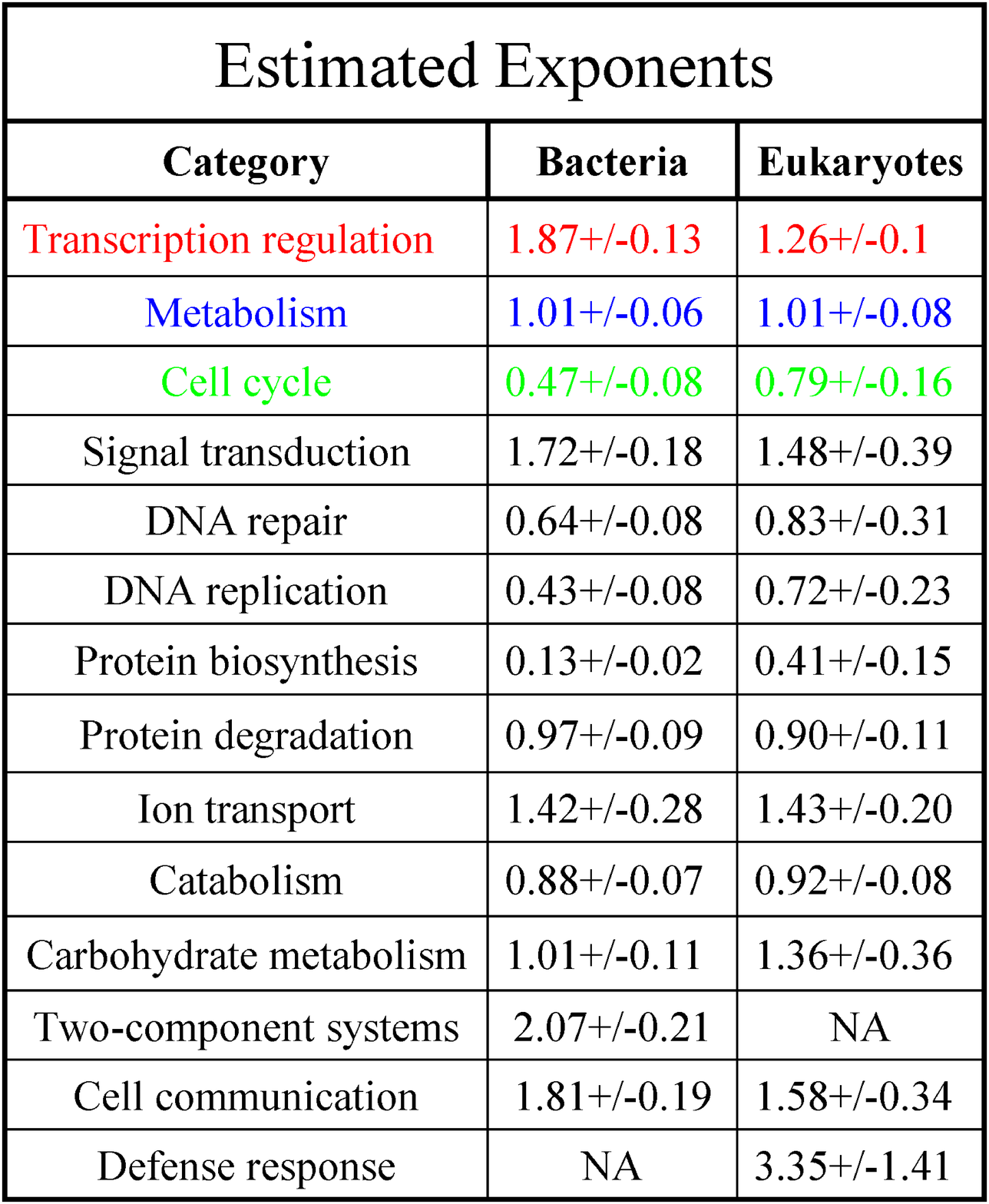,height=8cm}}
\caption{Estimates for the exponents of a selection of functional
categories. The first number gives the maximum likelihood estimate of
the exponent while the second indicates the boundaries of the $99$\%
posterior probability interval.}
\label{estimated_exponent_table}
\end{table}

The power-law fits and $99$\% posterior probability estimates for
their exponents were obtained using a Bayesian procedure described in
the appendix. To assess the quality of the fits, I measured, for each
fit, the fraction of the variance in the data that is explained by the
fit. In bacteria, $26$ out of $44$ GO categories have more than $95$\%
of the variance explained by the fit, $38$ categories have more than
$90$\% explained. In eukaryotes, $26$ categories have more than $95$\%
percent explained by the fit and $32$ have more than $90$\% explained
by the fit. However, with total gene number varying over less than two
decades in bacteria, and the small number of data points in
eukaryotes, one may wonder how one can claim that power-laws have been
``observed''. First, the fact that I fit the data to power-laws should
not be mistaken for a claim that the data can {\em only} be described
by power-law functions. I only claim that the power-law is by far the
simplest functional form that fits almost all the observed
data. Second, when scatter plots such as those shown in
Fig. \ref{scaling_fig} are plotted on linear as opposed to logarithmic
scales, it is clear even by eye that the fluctuations in the number of
genes in a category scale with the total number of genes in the
genome. That is, the fluctuations in the data suggest that logarithmic
scales are the ``natural'' scales for this data. This is further
supported by the simple evolutionary model presented below.

One may also wonder to what extent the results are sensitive to the
specific functional annotation procedure. I performed a variety of
tests to assess the robustness of the results, i.e. the observed
power-law scaling and the values of the exponents, to changes in the
annotation methodology (see appendix). These involve using entirely
independent annotations based on Clusters of Orthologous Groups of
proteins (COG) \cite{COGpaper97}, and a simple (crude) annotation
scheme based on keyword searches of protein tables for sequenced
microbial genomes from the NCBI website \cite{ncbi_ftp_website}. As
shown in the appendix, the observed power-law scaling, and the values
of the exponents are generally insensitive to these and other changes
in annotation methodology. It has to be noted, however, that all
currently available annotation schemes, including the ones used here,
predict function from sequence homology and thus at some level assume
that functional homology can be inferred from sequence homology. The
results reported here thus also depend on this assumption.

\section*{Observed Exponents}

Some functional categories, such as the large category of metabolic
genes, occupy a roughly constant fraction of a genome's gene-content,
as evidenced by their exponent of $1$. However, many categories show
significant deviations from this ``trivial'' exponent. Genes related
to cell-cycle or protein biosynthesis have exponents significantly
below $1$, whereas for transcription factors (TFs) the exponent is
significantly above $1$. These trends are strongest in bacteria. There
the exponent for TFs is almost $2$, implying that as the number of
genes in the genome doubles, the number of TFs quadruples. This has
some interesting implications for ``regulatory design'' in
bacteria. It implies that the number of TFs {\em per gene} grows in
proportion to the size of the genome (see
\cite{pseudomonas_genome_paper} for a similar observation). This in
turn implies that, in larger genomes, each gene must be regulated by a
larger number of TFs and/or each TF must be regulating a smaller set
of genes. An exponent of $2$ is also observed for two-component
systems\footnote{Note that two-component systems were not in the list
of $44$ high-level categories.}, which are the primary means by which
bacteria sense their environment. This suggests that the relative
increase in transcription regulators in more complex bacteria is
accompanied by an equal relative increase in sensory systems.

The difficulties with gene prediction and annotation in eukaryotes,
the small number of available genomes, and our lack of understanding
of the role of alternative splicing across eukaryotic genomes make it
premature to draw many conclusions from Fig. 1B. However, the main
trends from Fig. 1A are reproduced: the super linear scaling of TFs,
the sub linear scaling of cell-cycle genes, and the small exponents
for DNA replication and protein biosynthesis genes. 

The observed super linear scaling of TFs also has implications for our
understanding of ``combinatorial control'' in transcription
regulation. It is well established that in complex organisms,
different TFs combine into complexes to affect transcription
control. Therefore, a relatively small number of TFs can implement a
combinatorially large number of different transcription regulatory
``states'', which may correspond to particular external environments,
developmental stages, tissues, combinations of external stimuli,
etcetera. Each such regulatory state will be associated with a
unique set of genes that are expressed in that state. If the number of
such regulatory states were proportional to the total number of genes,
then the number of TFs would increase much more {\em slowly} than the
total number of genes. However, the scaling results show that,
instead, the number of TFs increases more rapidly than the total
number of genes. This thus implies that the number of regulatory
states is also ``combinatorial'' in the total number of genes: a
relatively small number of genes is used in different combinations to
implement combinatorially many regulatory states.

The picture that emerges is not of TFs being used in different
combinations to implement the regulatory needs of individual
genes. But rather that, as one moves from simple to more complex
organisms, the number of regulatory states grows so much faster than
the total number of genes that, even with combinatorial control of
transcription, the number of TFs grows much faster than the total
number genes.

\section*{Evolutionary model}

One of course wonders about the origins of these scaling laws in
genome organization, and I like to present some speculations in this
regard. Assume that most changes in the number of genes $n_c$ in a
functional category $c$ are caused by duplications and
deletions. Then, $n_c(t)$ generally evolves according to the equation
\begin{equation}
\label{nc_equation}
\frac{d n_c(t)}{dt} = (\beta(t) -\delta(t)) n_c(t) = \rho(t) n_c(t),
\end{equation}
with $\beta(t)$ and $\delta(t)$ respectively the duplication- and
deletion-rate of the genes in this category at time $t$ in the
evolutionary history of the genome. For simplicity of notation I have
introduced the difference of duplication and deletion rates $\rho(t)$,
which can be thought of as an ``effective'' duplication rate. This
rate $\rho(t)$ is presumably proportional to the difference between
the average probability that selection will favor fixation of a
duplicated gene from this category and the average probability that
selection will favor deletion of a gene from this category. Similarly,
the total number of genes $g(t)$ obeys the equation
\begin{equation}
\label{g_equation}
\frac{d g(t)}{dt} = \gamma(t) g(t),
\end{equation}
with $\gamma(t)$ the overall effective rate of gene duplication in the
genome at time $t$ in its evolutionary history. When we solve for
$n_c$ as a function of $g$ we find
\begin{equation}
n_c = \lambda g^{\langle \rho \rangle/\langle \gamma \rangle},
\end{equation}
where $\langle \rho \rangle$ and $\langle \gamma \rangle$ are the mean
effective duplication rates of genes in category $c$ and the entire
genome respectively, {\em averaged} over the evolutionary history of
the genome, and $\lambda$ is a constant that depends on the boundary
conditions. In order for all bacterial genomes to obey the same
functional relation, the constant $\lambda$ and the ratios $\langle
\rho \rangle/\langle \gamma \rangle$ have to be the {\em same} for all
bacterial evolutionary lineages. Since all life shares a common
ancestor, the boundary conditions for equations (\ref{nc_equation})
and (\ref{g_equation}) are trivially the same for all bacterial
lineages, implying that the constant $\lambda$ is indeed the same for
all bacterial lineages. In summary, simply assuming that changes in
gene-number occur mostly through duplications and deletions implies
our observed power-law scaling {\em if} the ratios $\langle \rho
\rangle/\langle \gamma \rangle$ are the same for all evolutionary
lineages.

I thus propose that the explanation for the observed scaling-laws is
that the ratios $\langle \rho \rangle/\langle \gamma \rangle$ are
indeed the same for all bacterial lineages, i.e. these ratios of
average duplication rates are ``universal constants'' of genome
evolution. For instance, the exponent $2$ for TFs in bacteria
indicates that, in all bacterial lineages, evolution selects
duplicated TFs twice as frequently as duplicated genes in general. It
seems likely that such universal constants are intimately connected to
fundamental design principles of the evolutionary process. It is
tempting to become even more speculative in this regard, and suggest
that this factor of $2$ in duplication rate is related to switch-like
function of transcription factors: with each addition of a
transcription factor the number of transcription-regulatory states of
the cell doubles. It is not entirely implausible to assume that with
twice the number of internal states available, the probability of such
a duplication being fixed in evolution is twice as large as the
probability of fixing a duplicated gene that does not double the
number of internal states of the cell.

\begin{figure}[htbp]
\centerline{\epsfig{file=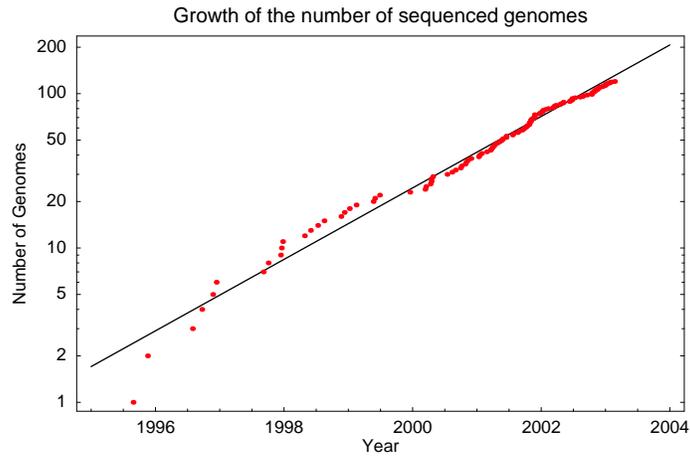,height=6cm}}
\caption{The number of fully-sequenced genomes in the NCBI database
\cite{ncbi_genomes} as a function of time. The vertical axis is shown
on a logarithmic scale. The straight line is a least squares fit to an
exponential function: $n(t)= 2^{(t-1994)/1.3}$.}
\label{genome_number_fig}
\end{figure}

Finally, as table \ref{estimated_exponent_table} shows, there is still
substantial uncertainty about the exact numerical values of the
exponents given the current data, and many more genomes are needed to
estimate these values more accurately. A survey of the NCBI genome
database \cite{ncbi_genomes} shows that the number of sequenced
genomes is increasing exponentially, with a doubling time of about
$16$ months (Fig. \ref{genome_number_fig}).
This suggests that within a few years thousands of genomes will become
available. With such an increase in available data it will become
possible to look at much more fine-grained gene content statistics
than the ones presented here. One can for instance imagine going
beyond looking at single functional categories at a time, and
investigate if there are correlations in the variations of gene number
in more fine-grained functional categories. I believe that such
investigations have the potential to teach us much about the
functional design principles of the evolutionary process.

\bibliography{/home/golem/erik/epev}
\bibliographystyle{plain}

\newpage 

\appendix

\section*{Appendix}

\subsection*{Results for Archaea}

\begin{figure}[htbp]
\centerline{\epsfig{file=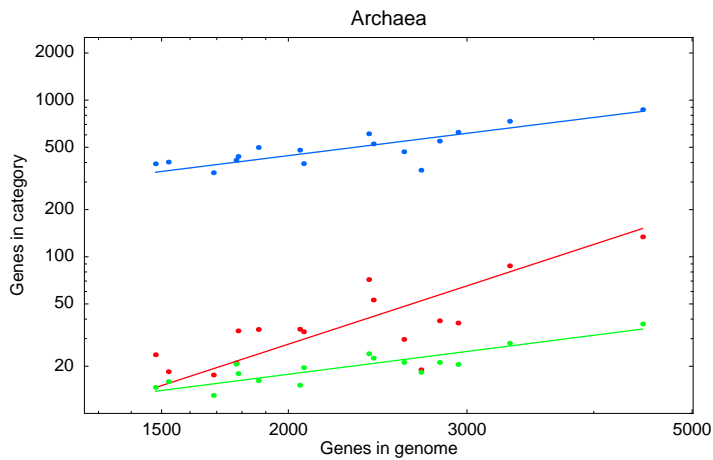,height=6cm}}
\caption{The number of transcription regulatory genes (red), metabolic
genes (blue), and cell cycle related genes (green) as a function of
the total number of genes in archaea. Both axes are shown on
logarithmic scales. Each dot corresponds to a genome. The straight
lines are power-law fits.}
\label{archaea_fig}
\end{figure}

Figure \ref{archaea_fig} shows the number of transcription regulatory
genes (red), metabolic genes (blue), and cell-cycle related genes
(green) as a function of the total number of genes in archaeal
genomes.
Note that the size of the largest archaeal genome differs from that of
the smallest by only a factor of approximately $3$. Consequently,
there is large uncertainty regarding the values of the exponents for
archaea. The maximum likelihood values and $99$\% posterior
probability intervals are $2.1$ and $[1.3-5.7]$ for transcription
regulatory genes, $0.81$ and $[0.44-1.48]$ for metabolic genes, and
$0.83$ and $[0.54-1.35]$ for cell-cycle related genes.

\subsection*{Power-law Fitting}

The power-law fits in figure \ref{scaling_fig} were obtained using a
Bayesian straight-line fitting procedure. For each GO-category, I
log-transform the data such that each data point $(x_i,y_i)$
corresponds to the logarithm of the estimated total number of genes,
and the logarithm of the estimated number of genes in the category
respectively. I assume that these transformed data where drawn from a
linear model
$$
y = \alpha (x + \epsilon) + \lambda + \eta,
$$
where $\eta$ and $\epsilon$ are ``noise'' terms in the $x$- and
$y$-coordinates respectively and $\lambda$ in an unknown off-set. I
assume that the joint-distribution $P(\eta,\epsilon)$ is a
two-dimensional Gaussian with means zero and unknown variances and
co-variance. That is, I use scale invariant priors for the variances,
and will integrate these nuisance variables out of the likelihood. A
uniform prior is used for the location parameter $\lambda$, and for
the slope $\alpha$ I will use a rotation invariant prior:
$$
P(\alpha)d\alpha = \frac{d\alpha}{(1+\alpha^2)^{3/2}}.
$$
The use of these priors guarantees that the results are invariant
under all shifts and rotations of the plane.

Integrating over all variables except for $\alpha$, we obtain
for the posterior $P(\alpha|D)$ given the data $D$:
$$
P(\alpha|D) d\alpha = C \frac{(\alpha^2+1)^{(n-3)/2}d\alpha}{(\alpha^2 s_{xx}-2
\alpha s_{yx} + s_{yy})^{(n-1)/2}},
$$
where $n$ is the number of genomes in the data, $s_{xx}$ is the
variance in $x$-values (logarithms of the total gene numbers),
$s_{yy}$ the variance in $y$-values (logarithms of the number of genes
in the category), $s_{yx}$ is the co-variance, and $C$ is a
normalizing constant. The values of the exponents reported in table
\ref{estimated_exponent_table} are the values of $\alpha$ that
maximize $P(\alpha|D)$, and the boundaries of the $99$\% posterior
probability interval around it. 

For each fit, I also measure what fraction of the variance in the data
is explained by the fit. That is, I compare the average distance $d$
of the points in the plane to their center of mass with the average
distance $d_l$ of the points to the fitted line and define the
fraction $q = 1-d_l/d$ as the variance in the data explained by the
fit.

\subsection*{Robustness of the results}

I first checked that the total amount of available annotation
information is not itself dependent on genome size. If the total
amount of available annotation information were to vary with the
number of genes in the genome, this could lead to biases in the
estimated exponents. To exclude this possibility, I counted the total
number of genes with {\em any} Interpro hits in each genome and found
that, for both bacteria and eukaryotes, the fraction of genes in the
genome with Interpro hits is about $2/3$, independent of the total
number of genes in the genome. Consistent with this observation, when
one fits power-laws to the number of genes $n_c$ in a GO-category $c$
as a function of the total number of {\em annotated} genes in the
genome, one finds exponents that are very close to those founds for
$n_c$ as a function of the total number of genes in the genome.

Second, I tested that the results are insensitive to the value of the
parameter $\beta$. The default value $\beta=3$ gives a gene with a
single Interpro hit a probability of $1-e^{-3} \approx 0.95$ to belong
to the category. This is reasonable because Interpro is designed to
only report statistically significant hits. To assess the effect of
changing $\beta$, results for $\beta=1$ were generated. For bacteria
the change in fitted exponent is less than $5$\% for $26$ of $44$
categories, and less than $10$\% for $39$ categories. For eukaryotes
the exponent changes by less than $5$\% for all but $3$ categories. In
all cases, the change in fitted exponent is significantly smaller than
the $99$\% posterior probability intervals associated with the
exponents of the fits.

Third, I tested the robustness of the results against removal of
potential ``redundancies'' in the data. For bacteria there are several
examples where multiple genomes of the same species or genomes of very
closely related species occur in the data, and one might suspect that
these may bias the results in some way. To this end, I parsed the
names of all bacterial species into a general and a specific part,
e.g. for {\em Escherichia Coli} {\em Escherichia} is the general part
and {\em coli} the specific part, for {\em Listeria innocua} {\em
Listeria} is the general part and {\em innocua} the specific part,
etcetera. Groups of genomes with the same general part were then
collected together and for each group the gene numbers were replaced
with a single average of total gene number and average gene counts in
each of the functional categories. This reduces the size of the data
set by about a third. Power-law fitting was then applied to this
reduced set and the fitted exponents were compared with those of the
full data set. The maximum likelihood exponent was changed by less
than $5$\% in $31$ out of $44$ categories. The largest observed change
was an $18$\% change in the exponent. All changes to the exponents
were well within the $99$\% posterior probability intervals.

More importantly, the results could depend on the use of Interpro, the
Gene Ontology, and the mapping of Interpro entries to GO
categories. To test the robustness of the results to biases inherent
in Interpro annotation and/or the mapping from Interpro to Gene
Ontology, I analyzed selected functional categories using two other
annotation schemes.

The first is based on Clusters of Orthologous Groups of proteins (COG)
\cite{COGpaper97} annotation of $63$ bacterial genomes that can be
obtained from\\ ftp://ftp.ncbi.nih.gov/pub/COG/COG.\\ In this data
set, proteins of the $63$ bacterial genomes are assigned to COGs, and
the COGs have been assigned to functional categories. I used these
assignments to count the number of genes in different functional
categories according to the COG annotation scheme. A comparison of the
exponents for COG functional categories and the exponents for the
closest GO categories obtained using the Interpro annotations are
shown in table \ref{comparison_table}.

The second alternative annotation scheme I used is based on simple
keyword searches of protein tables for fully-sequenced bacterial
genomes available from the NCBI ftp site:\\
ftp.ncbi.nlm.nih.gov/genomes/Bacteria.\\ Removing genomes for which
little or no annotation exists, this leaves protein tables for $90$
bacterial genomes. Each protein in these protein tables is annotated
with a short description line. The number of genes in different
functional categories was counted by searching each description line
for hits to a set of keywords that characterize the category. For
instance, I chose the keywords ``ribosom'', ``translation'', and
``tRNA'' for the category ``protein biosynthesis'', and the gene is
counted as belonging to this category if any of these keywords occurs
in its description line. For the other categories I used the following
keywords: ``transcription'' for the category ``transcription'';
``transport'', ``channel'', ``efflux'', ``pump'', ``porin'',
``export'', ``permease'', ``symport'', ``transloca'', and ``PTS'' for
the category ``transport''; all combinations ``X Y'' with X being one
of ``ion'', ``sodium'', ``calcium'', ``potassium'', ``magnesium'', and
``manganese'', and Y being one of ``channel'', ``efflux'',
``transport'', and ``uptake'' for the category ``ion transport'';
`protease'' and ``peptidase'' for the category ``protein
degradation''; ``kinase'' for the category ``kinase''; and finally the
phrases `DNA polymerase'', ``topoisomerase'', ``DNA gyrase'', ``DNA
ligase'', ``replication'', ``helicase'', ``DNA primase'', ``DNA
repair'', ``cell division'', and ``septum'' for the category ``cell
cycle''. The exponents resulting from this (crude) annotation scheme
are also shown in table \ref{comparison_table}. As table
\ref{comparison_table} shows, there is good quantitative agreement
between the exponents that are obtained with the different annotation
schemes.

\begin{table}[htbp]
\caption{Estimated $99$\% posterior probability intervals for the
scaling exponents obtained with three different annotation schemes.}
\begin{center}
\begin{tabular}{|c|c|c|} \cline{1-3}
Annotation &  Category & Exponent\\
\hline
\hline
GO & Protein biosynthesis & 0.11-0.15\\
COG & Translation, ribosomal structure and biogenesis & 0.21-0.37 \\
NCBI & Protein biosynthesis & 0.09-0.15\\
\hline
GO & Signal transduction & 1.55-1.9\\
COG & Signal transduction mechanisms & 1.66-2.14\\
\hline
GO &  Protein metabolism and modification & 0.68-0.8 \\
GO & Protein degradation & 0.89-1.06\\
COG &  Posttranslational modification, protein turnover, chaperones &
0.88-1.15\\
NCBI & Protein degradation & 0.68-0.91\\
\hline 
GO & Cell cycle & 0.39-0.54\\
GO & DNA repair & 0.52-1.14\\
COG & Replication, recombination and repair & 0.66-0.83\\
NCBI & Cell cycle & 0.45-0.64\\
\hline
GO & Ion transport & 1.15-1.7\\
COG & Inorganic ion transport and metabolism & 1.19-1.47\\
NCBI & Ion transport & 1.12-1.88\\
\hline
GO & Regulation of transcription & 1.74-2\\
COG & Transcription regulation & 1.69-2.36\\
NCBI & Transcription & 1.9-2.42\\
\hline
GO & Kinase & 0.96-1.16\\
NCBI & Kinase & 0.8-1.03\\
\hline
GO & Transport & 1.08-1.32\\
NCBI & Transport & 1.16-1.5\\
\hline
\end{tabular}
\end{center}
\label{comparison_table}
\end{table}

\end{document}